\documentclass[twocolumn,twoside,slac]{revtex4}
\usepackage{graphicx}
\usepackage{fancyhdr}
\begin{document}

\title{EU DataGRID testbed management and support at CERN}

\author{E. Leonardi}
\affiliation{CERN, Geneva, Switzerland}
\author{M.W. Schulz}
\affiliation{CERN, Geneva, Switzerland}

\begin{abstract}

In this paper we report on the first two years of running the CERN testbed
site for the EU DataGRID project. The site consists of about 120 dual-processor
PCs distributed over several testbeds used for different purposes: software
development, system integration, and application tests. Activities at the site
included test productions of MonteCarlo data for LHC experiments, tutorials and
demonstrations of GRID technologies, and support for individual users analysis.
This paper focuses on node installation and configuration techniques,
service management, user support in a gridified environment, and includes
considerations on scalability and security issues and comparisons with
"traditional" production systems, as seen from the administrator point of view.

\end{abstract}

\maketitle

\thispagestyle{fancy}

\section{\label{Intro}Introduction}

The EU DataGRID (EDG) project \cite{EDG} started in January 2001 and is now
entering its third and last year of activity. The goal of the project is the
development of a consistent set of services to create a distributed computing
grid, as defined in \cite{TheGRID}.To test the system under development, a
distributed testbed was created involving at first the five main partners of
the project (CERN, INFN, IN2P3, PPARC, and NIKHEF). This testbed was then
extended to more and more sites, to a total which now exceeds 20 centers all
over Europe.

Since the beginning of the project, CERN has been in a very central position,
from development to final deployment of the software on the main testbed.
In this paper we will describe our experience as managers of the CERN testbed
site.

\section{\label{EDGServ}Description of EDG services}

The current (as of March 2003) release of the EDG middle-ware includes a
set of services which implement, even if in a non-definitive way, all the
basic services needed to create a grid. In this section we will give an
overview of these services.

\subsection{Authentication}

User and node authentication complies with the Grid Security Infrastructure
(GSI), as defined in \cite{GSI}, and is based on the openSSL implementation
of the Public Key Infrastructure (PKI).

\subsection{Authorization}

Access to resources is authorized through a simple map file which associates
each authorized user certificate with a dynamically assigned local account.

The list of authorized certificates is available from a small number of LDAP
servers, each managed by a different Virtual Organization (VO). As a
consequence, the current authorization structure is very coarse grained and
can only distinguish either between two individual users or between users
belonging to different VO's.

\subsection{Resource Access}

Physical access to local resources, i.e. submission of jobs to a local batch
system or transfer of a file to/from a local storage server, uses the
gatekeeper service provided by the GLOBUS \cite{GLOBUS} project with the
addition of EDG specific extensions, and the GSI-enabled GridFTP server, also
from GLOBUS.

\subsection{Storage Management}

Storage management is still at a quite early stage of development and includes
a file replication service (GDMP \cite{GDMP}), which at CERN is interfaced to
CASTOR, the local MSS service \cite{CASTOR}, and an LDAP-based Replica Catalog,
which will be described later in some detail.

\subsection{Information Services}

Information about available resources is distributed via a hierarchical Grid
Information Service (GIS), also using LDAP as the base protocol.

\subsection{Resource Management}

This is the intelligence of the grid: a Resource Broker (RB) scans job
requirements, chooses the best matching site using information from the GIS
and delivers the job to the corresponding gatekeeper.

To accomplish this task, the top RB server uses several subordinate services:
Job Submission Server (JSS) Job Parsing Server (JPS), CondorG \cite{CondorG},
...

\subsection{Logging and Bookkeeping}

The Logging and Bookkeeping (LB) service stores a state-transition based
history of each job handled by the grid.

\subsection{Monitoring and Accounting}

Monitoring of the correct functioning of each EDG service and of the grid as a
whole is foreseen for the final version of the EDG software but no integrated
monitoring system was available at the time of the report. The same applies for
a detailed accounting of resource utilization.

{\sl Ad hoc} solutions for both services were in operation and varied from site
to site.

\subsection{Node Installation}

One of the goals of the EDG project is to provide automatic tools to manage
large computing centers in the LHC era. Most testbed sites used an
automatic node installation and management tool, the Local Configuration System
(LCFG \cite{LCFG}), which will be described later in this paper.

\section{\label{CERNTestbed}EDG testbeds at CERN}

Due to its central position within the project, CERN hosts several parallel
testbeds, used for different tasks.

\subsection{Application testbed}

On this testbed the latest ``stable'' version of the EDG middle-ware is
installed. This testbed is open for all users participating to the project and
is used for several different tasks:

\begin{itemize}
\item programmed ``stress tests'' to evaluate the behavior of the middle-ware
on a medium scale distributed testbed;
\item data production tests by all application groups, including all LHC
experiments, bio-medical applications, and earth observation projects;
\item demonstrations and tutorials of EDG middle-ware functionalities.
Incidentally we can note that these activities, targeted at increasing the
public visibility of the project, often posed problems to the site managers:
given the current maturity level of the software, to avoid putting the
demonstration at risk a constant surveillance of the testbed was required for
the whole duration of the same and no other activities (test, normal usage)
could take place. Recently an independent testbed, specific for demonstrations,
has been set up.
\end{itemize}

The number of sites participating to this testbed grew from the original 5
sites to the current 20 and is constantly growing. CERN hosted most of the
central services (the top node of the information system, Replica Catalogs
for two VO's, two of the main RB's, ...) and connected to the middle-ware a
PBS-based batch system with a number of Worker Nodes which went up to over 100
in periods of intense activity.

As only software versions which passed a preliminary extensive test were
installed here, this testbed underwent a relatively small number of software
updates, mostly required for security patches, installation of new applications
on behalf of the experimental groups, and modifications to the access
permissions. On the other hand, being aimed at production tests, this
testbed required a fairly large amount of manpower due to the instability of
the software: most of the main services needed a complete restart at least once
per day and, at the same time, a lack of management tools and experience made
troubleshooting very hard.

\subsection{Development testbed}

On this testbed all EDG software versions are integrated and tested. In order
to keep response and update time as short as possible, only the five main sites
participate to this testbed.

Aimed at functionality testing, the number of nodes in this testbed was fairly
small compared to the application testbed. At CERN only up to 10 nodes were
used for the development testbed.

Update activity on this testbed was continuous, often requiring installation of
several versions in a single day. To facilitate troubleshooting,
developers had direct access to most service nodes. On several occasions this
induced some traceability problems as it was not always easy to get a complete
description of what the developer did to fix a bug, especially if this included
node configuration changes.

\subsection{Integration testbed}

Last April, a new testbed was created to perform the (still ongoing)
integration of version 2.0 of the EDG middle-ware.

This testbed is composed of about 20 nodes and has characteristics very similar
to the development testbed: continuous deployment of test versions, free access
to developers, frequent re-installation of the nodes.

On this testbed, the EDG middle-ware was first ported to the RedHat 7.3 version
of Linux and to the 2.2.4 version of the GLOBUS toolkit, in its VDT 1.1.8
\cite{VDT} manifestation.

Since the integration phase for EDG 2 started, most of the EDG manpower and
hardware resources at CERN were diverted to this testbed so that, at the time
of writing, the CERN development testbed has been completely dismantled and the
application testbed has been reduced to just a Storage Element (SE) node
providing access to the CASTOR MSS.

\subsection{Developers' testbeds}

Before a new version of one of the services is allowed to be merged with the
rest of the middle-ware on the development or integration testbed, it must be
tested for basic functionality in standalone mode. To this end we set up
several ``reduced'' testbeds, including only software components needed to test
one of the services at a time.

Each of these testbeds consisted of only two or three nodes but there were
continuous requests for the creation of topologies centered on any one of the
services.

\subsection{Testbed infrastructure}

To support all these testbeds, CERN provided an infrastructure based on some
standard CERN computer center services with the addition of a few EDG specific
services:

\begin{itemize}
\item 5 data servers, with a total of 2.5 TB of mirrored disk storage, offered
NFS-based storage capabilities for EDG users' home directories, shared areas
for the batch system (the gass\_cache area described later) and generic storage
for data production tests;
\item one of the EDG SE's was interfaced to the CASTOR MSS;
\item as many of the EDG developers and users could not be provided with a
CERN AFS account, we set up a NIS service for all EDG users;
\item a Certification Authority (CA) \cite{CERNCA} was set up specifically for
EDG usage and provided personal and host certificates for CERN users. This CA
is now evolving toward a CERN-wide service.
\item an LCFG server, including a private DHCP server, was setup to support
node installation and configuration.
\end{itemize}

The whole testbed infrastructure was then interconnected to the standard
CERN LAN with 100 Mbps or, in the case of data servers, 1 Gbps Ethernet lines.

\section{\label{EDGRel}EDG releases}

Before March 2002, no real software release procedure was in place. This led to
several problems mostly related to the tracing of hand-made modifications on
the configuration of the nodes and the installation of private executables by
the developers. In turn, this resulted in misalignment among the five
participating sites and in huge difficulties in collecting a set of software
packages and configuration settings to create a consistent version of the EDG
middle-ware.

In spite of the difficulties and with a lot of hard work from all the people
involved, we were able to converge to the first real release of the EDG
software, version 1.1.2, which was then used in some key demonstrations of
functionalities in March 2002.

To improve the situation, a strict release procedure was defined:

\begin{itemize}
\item all new rpms are first delivered to the Integration Team which takes care
of inserting them in a new CVS-based tentative release tag;
\item this tentative tag is first installed on the CERN development testbed
and a predefined set of basic tests is applied;
\item the five core sites install the same tag and a new set of tests, centered
on distributed functionalities, is applied;
\item the tag is installed on the application testbed where scalability tests
and generalized use can begin.
\end{itemize}

If at any of these stages the tests fail, the tag is rejected and bug reports
are sent back to the relevant developers.

To improve flexibility, application software only needed at the final stage
of testing to create a realistic environment is not required to follow this
procedure and can be installed upon request of the main application groups.
Also, basic OS security patches and support for new CA's can be applied at
need.

Thanks to this procedure, the release of new versions of the code proceeded
in a much smoother way and had its finest day last November when all EDG
sites successfully moved from version 1.2.3 to the non-downward-compatible
version 1.3.0 in only one day.

\section{\label{NodeInst}Node installation}

One of the key issue to implement the release procedure described in the
previous section is the possibility of installing and configuring service nodes
located at geographically remote sites according to a predefined and
continuously changing set of instructions.

To this end EDG adopted and extended the Local Configuration System (LCFG)
developed at the University of Edinburgh.

This tool uses a human-readable description of the basic configuration of the
Linux OS and sends it in XML format to an agent on the node for installation.
There a set of scripts, called ``objects'', use this description to actually
modify the system configuration files.

LCFG can easily be extended by providing additional objects which can configure
any new functionality or service one may want to add.

After a slow start, more and more objects were created to effectively configure
all EDG specific services. Today only very few hand configuration steps are
still needed to create an EDG-enabled site and even these will be soon
completely automated.

Even if LCFG has proved to be the most valuable instrument to keep the EDG
release procedure in track, we found a few drawbacks which had to be taken into
account before adopting this tool:

\begin{itemize}
\item new objects have to be carefully designed in order not to interfere with
other objects;
\item some basic limitations of the standard objects have to be taken into
account, e.g. hard disks can only be partitioned using the four primary
partitions;
\item no feedback about the installation or update process is sent back to the
LCFG server: this required the creation of {\sl ad hoc} tools to do some
basic checks and a lot of manual work in case of problems;
\item LCFG wants to have total control of the machine configuration, from
basic OS functions to application level. This means that, in its basic
version, LCFG is not suited to install, for example, the EDG middle-ware on top
of an already installed node. Recently a modified version of LCFG, called
LCFGlite\cite{LCFGlite}, was developed to handle this case;
\item due to its structure, LCFG does not cope well with OS settings which may
change outside of its control. An example are user passwords: if users change
their passwords, LCFG will change them back to a predefined value. To solve
this problem, we moved all user accounts to a non-LCFG-managed NIS server and
left only system accounts under local LCFG control;
\item as each modification to the node configuration must be first inserted
into the configuration description file on the main server, using LCFG to
manage nodes used by developers for their first tests might substantially slow
down the fast rate of changes. Also, if a developer modifies any
configuration file by hand, this might be changed back by LCFG, thus
introducing a lot of entropy into the process.
\end{itemize}

To overcome part of these shortcomings, the EDG WP4 group is currently
finalizing an alternative tool which will replace LCFG.

In parallel with LCFG, we used the network boot capabilities of recent NIC's
and the syslinux tool \cite{syslinux} to bootstrap the initial installation
procedure of the nodes directly from the network. It was then sufficient to
start a private DHCP server on the LCFG server to get the whole node
installation and configuration automatically accomplished.

To improve the situation even further, we implemented a system to remotely
reset nodes using a serial line controlled relay system connected to the
reset switch of the motherboards and we used multi serial line boards to
collect all consoles to a few central servers, thus allowing a completely
web-based control of each node. A report on this project was presented at this
conference in \cite{Andras}.

After all the described tools were in place and after an initial period of
adaptation and tuning, the large number of rapidly changing nodes needed for
the EDG deployment became much more manageable and our visits to the CERN
computer center decreased to almost nil.

\section{\label{Middleware}Middle-ware}

The complexity and novelty of the EDG project made the list of problems
encountered in the integration and testing phases particularly long. Here we
briefly list some of the main issues which emerged in the process and describe
in more detail the problems which have interesting aspects from
the system- and site-manager point of view.

As noted previously, many of the middle-ware services were and still are quite
fragile, needing frequent restarts. In addition to this, the fault patterns
tend to be very complex, often involving several services at the same time,
thus making the troubleshooting process quite hard and the learning curve
very steep. This state of affairs, normal for an R\&D project of this size, was
worsened by the fact that the overall architectural design of the project
concentrated on defining service functionalities and their interaction but
often neglected to deal with the resource management aspects of the problem.

As a consequence, we had many problems in dealing with otherwise normal aspects
of storage management: adding new disk space was very tricky, as well as moving
file around within the available disk space. Also, no tools to handle scratch
space on either batch or service nodes were foreseen, and it was very easy
to see a whole disk fill up bringing the node, and often with it the whole
grid, to a grinding halt. Log files, one of the main tools to perform
troubleshooting and system checking, were most of the time hard to interpret,
often lacked some fundamental information (the time tag!), and for a few
services did not exist at all.

Another aspect which was not sufficiently taken into account in the first
development of the software was that of scalability: several services, even if
they correctly implemented all basic functionalities, could not scale beyond a
minimal configuration. Examples of this could be found in the Resource Broker,
which, due to a known limitation of the CondorG system, was not able to handle
more than 512 jobs at the same time, and the Replica Catalog, which had
problems if more than O(1000) files (this number depends on the average length
of the file name) were listed as a single collection. Even the Information
Service started malfunctioning as soon as the number of sites connected to the
grid increased from the initial five.

Parallel to scalability problems, we encountered problems related to more
traditional aspects of distributed system programming: all along the
development and integration process, besides the usual memory leak bugs, we
saw port leaks, file leaks, even i-node leaks. These problems were often
related to lower levels of the software and led to non obvious fault patterns
which needed a long time to debug.

A problem which haunted us in the early phases of the project was the lack of
control in the packaging process: executables which worked correctly when
compiled by hand by developers, suddenly started misbehaving when inserted
into a standard rpm package for deployment. This was most of the time due to
differences in libraries on private machines. To reduce this risk, an
auto-build system was created and is now in operation: developers are required
to set their packages so that they can be compiled on the official auto-build
node and only packages created by this system are accepted for release.

Before going into the details of some of the listed problems, we should note
that the experience from the first phase of the project was very valuable to
define the priorities for the new and final release of the EDG software,
currently in the integration phase, so that most of the problems and
inconsistencies found during the deployment of the first version of the
software were addressed and solved, thus improving stability, scalability, and
manageability of the whole system.

\subsection{The gass\_cache area}

Due to the internal structure of the GRAM protocol, defined by the GLOBUS
project to handle job submission and information passing in a grid-enabled
batch system, a disk area, known as the gass\_cache area, must be shared
between the gatekeeper (the borderline node between the grid and the
batch system) and all the worker nodes of the batch system.

According to the GRAM protocol, each job which is submitted to the
batch system creates a large number ($\gg 100$) of tiny files in the
gass\_cache area. If the job ends in an unclean way, very often these files are
not erased. In addition, due to a bug in the implementation, even
if the job ends correctly, the gass\_cache area is not completely cleaned.

Given the small size of the files, often not exceeding 100 bytes, these two
problems create a steady leak of i-nodes on the shared area so that, even if
this area appears to be almost empty, the GRAM protocol suddenly stops working
as no more files can be created. Being at the heart of the job submission
system of the grid, problems with the GRAM protocol manifest themselves in
a whole set of different and apparently uncorrelated malfunctions of
several subsystems: the first few times it took us a long time to find where
the problem lay.

Even knowing the real source of the problem, fixing it requires a long time
to clean up the huge number of leftover i-nodes, time during which the local
batch system is not visible from the grid, and the loss of any job running on
the batch system at the time.

As there is no easy way to map the internal structure of the gass\_cache area
to the jobs which are currently running on the batch system, and no tool to do
this was provided, either by GLOBUS, or by EDG, it is not possible to
create a clean up daemon which keeps an eye on the shared area and cleans up
files which are not longer needed.

\subsection{Storage management issues}

As already noted, the first version of the EDG middle-ware did not contain a
complete and integrated model for grid-wide storage management but only a set
of low level tools and services which could be used to partially interface
storage servers to the grid.

The principal tools were:

\begin{itemize}
\item a GSI-enabled ftp server;
\item GDMP \cite{GDMP}, a replication-on-demand service
\item a basic Replica Catalog (RC), with the cited limitations in the number of
files it can manage;
\item a set of user level commands to copy files to and from a data server,
to trigger file replication between data servers, and to register files to the
RC;
\item a basic interface to mass storage systems like CASTOR and HPSS.
\end{itemize}

A constraint in the way the RC was organized had unforeseen consequences on the
possibility of setting up and organizing a storage system which are worth
examining in some detail.

The RC is a simple database, in its first implementation based on the OpenLDAP
\cite{OpenLDAP} package, which contains a collection of logical file names
(LFN), used to uniquely identify a given file, and for each of them one or more
physical file names (PFN) which point to the physical location of each replica
of the logical file. By design, the PFN was obtained from the LFN by appending
it to the path of the grid-enabled area assigned to the corresponding VO on a
given storage system.

As an example, let's assume that a simulation job submitted by a physicist
belonging to the Atlas VO produces a file with LFN
{\tt prod/feb2003/simu001.fz}. If a copy of this file is stored on a SE at
CERN, it will have a PFN like
{\tt //lxshare0384.cern.ch/flatfiles/atlas/prod/}\\
{\tt feb2003/simu001.fz} where:

\begin{itemize}
\item {\tt lxshare0384.cern.ch} is the node name of the SE
\item {\tt /flatfiles/atlas} is the path to the disk area assigned to Atlas
\end{itemize}

This apparently harmless limitation had heavy implications in the usage of
storage systems: for the file replication system to work in a user transparent
way within the whole grid, the disk area on each SE must either consist of a
single big partition, which merges all physical disk systems available to the
node, or all SE's must have exactly the same disk partition structure.

To understand this, assume that at CERN the {\tt prod/feb2003} path corresponds
to a mount point of a 100 GB disk (the standard organization of CERN disk
servers) so that only Atlas can use it. At, e.g., RAL the {\tt prod/feb2003}
path might not exist yet, so if a user wants to replicate the file there he/she
must first create it. At RAL this path will end up on a partition which depends
on the disk layout of the local SE and which may very well not have enough
disk space to hold the replicated file, even if the SE itself has plenty of
free disk space on other partitions.

This problem was particularly annoying at CERN when we tried to set up a
central SE where the CMS collaboration could collect all simulated data
produced on the grid during its ``stress test''.
To allow for
automatic replication from the other sites, we had to carefully plan the
partition layout on the central SE, taking into account how much data would
be produced on average at each site participating to the test, and then ask
the CMS production group to store their data using different paths according
to which site the data were being produced at. The final system worked fine
but it was very far from the expected transparent replication of data within
the grid.

\subsection{Resource Broker issues}

As explained above, the Resource Broker (RB) is the central intelligence of the
grid, taking care of interpreting user requests expressed in the Job
Description Language (JDL) and mapping them to the available resources using
the Grid Information Service.

Due to its central role, the RB interacts with most of the grid services in
many different and complex ways. A malfunction in the RB is then very visible
from the user point of view as jobs are no longer accepted and it can have
bad effects on all services with which it interacts. A lot of effort was put
into fixing problems but it is still the most sensitive spot in the grid.

To improve reliability in job submission, several RB's have been set up at
different sites, thus increasing the scalability of the grid in terms of the
maximum number of jobs accepted and giving users some back-up access points to
the grid in case the local RB gets stuck with a problem.

Due to the large number of low-level services it uses, several problems can
show up in the RB, thus affecting the functioning of the whole grid. One of
these is the corruption of the job requests database. This is related to a
non-thread-safe low level library used by the postgres database which, on
dual-processor nodes, can lead to data corruption.

When this problem occurs, all RB daemons appear to run correctly but newly
submitted jobs end up in a waiting status. All the daemons must then be
stopped, the whole postgres database cleaned up, and then the RB can be
restarted. In the process, all active jobs being controlled by the RB
are lost. During normal usage of the system, this problem occurs on
average once per day per RB node.

It must be noted that a solution to this problem is already available and will
be deployed with version 2 of the EDG middle-ware.

\section{\label{Conclusions}Conclusions}

EDG testbeds have been in operation for almost two years, always providing
continuous and indispensable feed-back to EDG developers. LHC experiments
and the other project partners were able to get a first taste of a realistic
grid environment in preparation for future large scale deployments.

Being one of the most advanced grid projects currently in operation, most of
the problems and inadequacies of the EDG middle-ware were hard to isolate
and to fix, mostly due to lack of previous experience.

Several new problems related to resource management in a gridified environment
were isolated and are being or will be addressed in the final version of
the EDG software and related projects like LCG \cite{LCG}.

\begin{acknowledgments}
The authors wish to thank the EU and the national funding agencies for their
support of this work.
\end{acknowledgments}

\vfill

\end{document}